\documentclass{article}
\usepackage{hiph-preprint}
\usepackage{graphicx}
\usepackage{amssymb}
\volnumber{22} \issuenumber{1} \edyear{2005}                             
\frompage{000} \topage{000}                                              
\recrevdate{8 November 2005}                                              
\def\rightmark{$\Xi$ baryon correlations in $p+p$}
\def\leftmark{B.I.Bezverkhny}

\title{High Transverse Momentum $\Xi$ Baryon Correlations in p+p Collisions at $\sqrt{s}=200~GeV$.}
\authors{
{Betty I. Bezverkhny$^1$ for the STAR Collaboration %
\index{One, A.} 
\index{Two, A.} 
}\\[2.812mm]
{\normalsize \hspace*{-8pt}$^1$ Yale University, New Haven, CT
USA\\[0.2ex]
}}

\abstract{Angular correlations between produced high $p_T$ $\Xi^-$
baryons and unidentified charged and neutral particles are observed
in high-tower-triggered (on a large electromagnetic energy deposit)
$\sqrt s=200$ GeV $p+p$ collisions. This trigger favors events with
higher average multiplicity than those in minimum bias. These events
are likely to contain jets. The average multiplicity of a high-tower
triggered event is similar to that of a minimum bias event
containing a $\Xi$ baryon, which implies that $\Xi$ baryons are
likely to be produced in jets. $\Xi^-$ $\langle p_T\rangle$ is
higher in the triggered data then in the minimum bias sample.}


\makeindex
\begin{document}

\maketitle
\setcounter{page}{1}

\section{Introduction}\label{intro}
STAR has measured high-$p_T$ charged hadron suppression in the most
central Au+Au collisions via unidentified azimuthal correlations
\cite{jets} and the corresponding nuclear modification factors
\cite{raa}. Jet reconstruction can be difficult in a
high-multiplicity Au+Au collision environment, therefore
two-particle correlations are employed to study jets statistically.
As $p+p$ collisions are used as a reference for a Au+Au jet
measurement, the statistical method is applied to $p+p$ collisions
as well.

Identified strange high $p_T$ correlations are another step towards
furthering our understanding of the matter produced at RHIC, since
strange particles are created in all collision systems. Moreover,
strange particles provide a convenient comparison of produced quark
($\bar{s}s$) content vs. particle mass. They are also a tool in
understanding the baryon/meson differences, especially useful
because particle identification out to high $p_T$ is possible. Thus
strange particle correlations may yield new insights into the
strangeness production mechanism in both p+p and Au+Au collisions.

\section{Detectors and Data}\label{datadet}

\subsection{Detectors}\label{detectors}
The $2\pi$ azimuthal coverage of the STAR detector makes it
well-suited to study identified high-$p_T$ two-particle
correlations. The strange particles are reconstructed topologically
via their decay products in the STAR Time Projection Chamber (TPC)
\cite{topReco}, and correlated with charged tracks, likewise
reconstructed using the TPC \cite{bbHQ04}. We also construct
correlations between $\Xi$ candidates and energy deposits in the
STAR Barrel Electromagnetic Calorimeter (EMC), which are a result of
electromagnetic showers from both charged particles and photons
(either direct or $\pi^0$ daughters). In this analysis the EMC
covers $2\pi$ in azimuth and 1 unit of positive pseudorapidity. It
consists of three detectors: high energy towers (each having
dimensions of $\Delta\eta=\Delta\phi=0.05$) and two layers of
shower-max detectors, SMD-$\phi$ and SMD-$\eta$. Using energy
clusters from all three, an EMC point is reconstructed, which can
then be used in correlation analyses.

\subsection{Data}\label{data}
Although STAR recorded a 14 M minimum bias (min. bias) event sample
in 2002, less than a thousand $\Xi^-$ and $\overline{\Xi}^+$
candidates with $p_T>2$ GeV/c were reconstructed \cite{bbHQ04}. To
gain in statistics, we examine the 2 M events of the 2004 data
triggered on a large localized deposit of energy in the STAR EMC.
The events in this triggered data sample consist events triggered on
a minimum 2.5 GeV deposited in one EMC tower. Assuming that in a
$p+p$ collision high energy particles are more likely to be created
via parton fragmentation, this trigger selects data with more jets
than in a min. bias sample. The increase in jettiness appears to
correlate to increased mean multiplicity of the events \cite{gans}.
When an event is triggered by a deposit above threshold in one of
the towers, data from other towers are also recorded.

\section{Analysis and Results}\label{analysis}
\begin{figure}[t!]
\begin{minipage}[t!]{2.3 in}
\vspace*{-0.5cm}
  \includegraphics[width=2.15in]{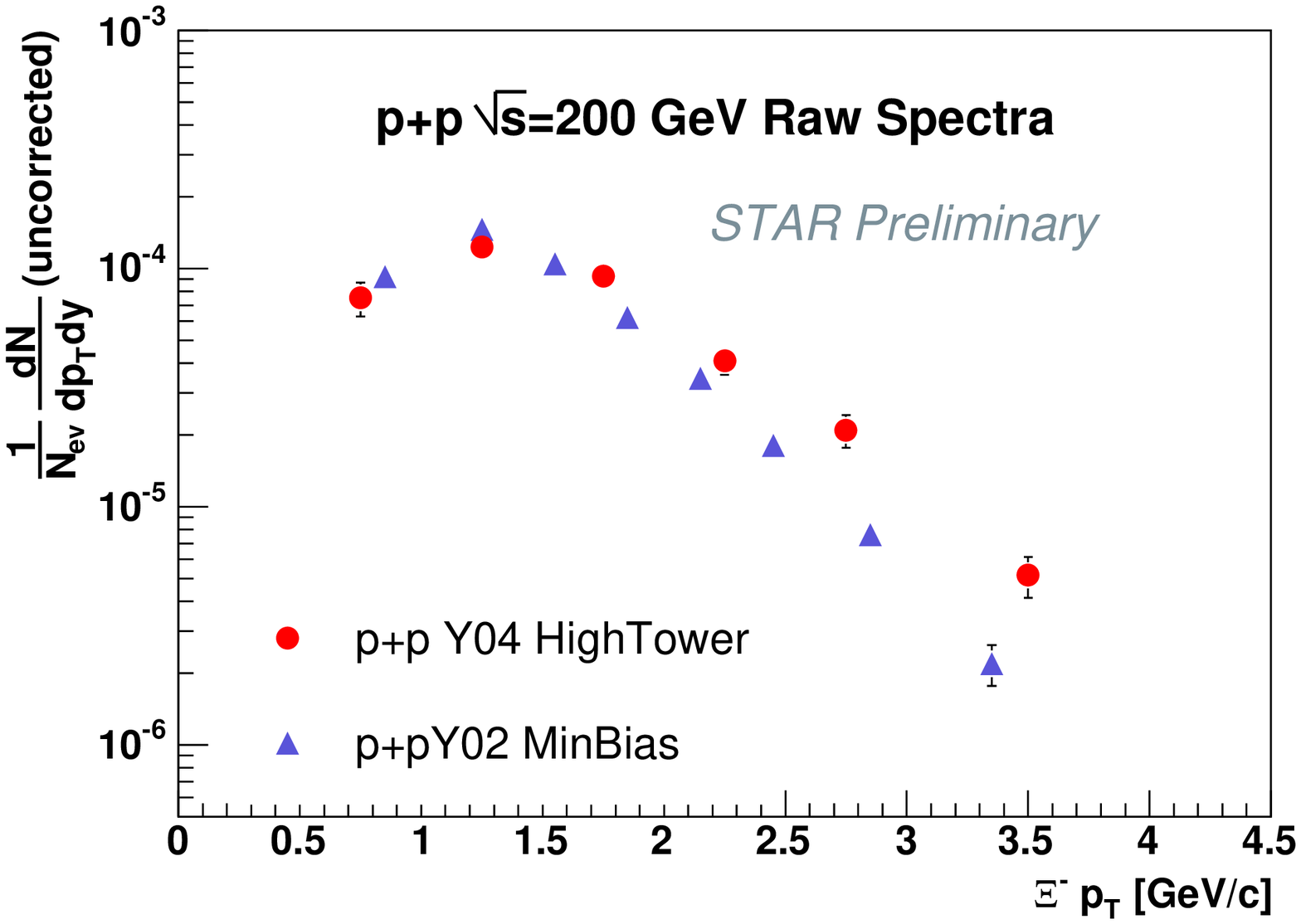}\\
\vspace*{-1.4cm}\\
  \caption{Uncorrected $\Xi^-$ spectra in minimum bias (triangles) and
  high tower (circles) events from $\sqrt s=200$ GeV $p+p$ collisions.}\label{fig1}
\end{minipage}
\hfill
\begin{minipage}[]{2.3 in}
\vspace*{-.2cm}
 \includegraphics[width=2.2in]{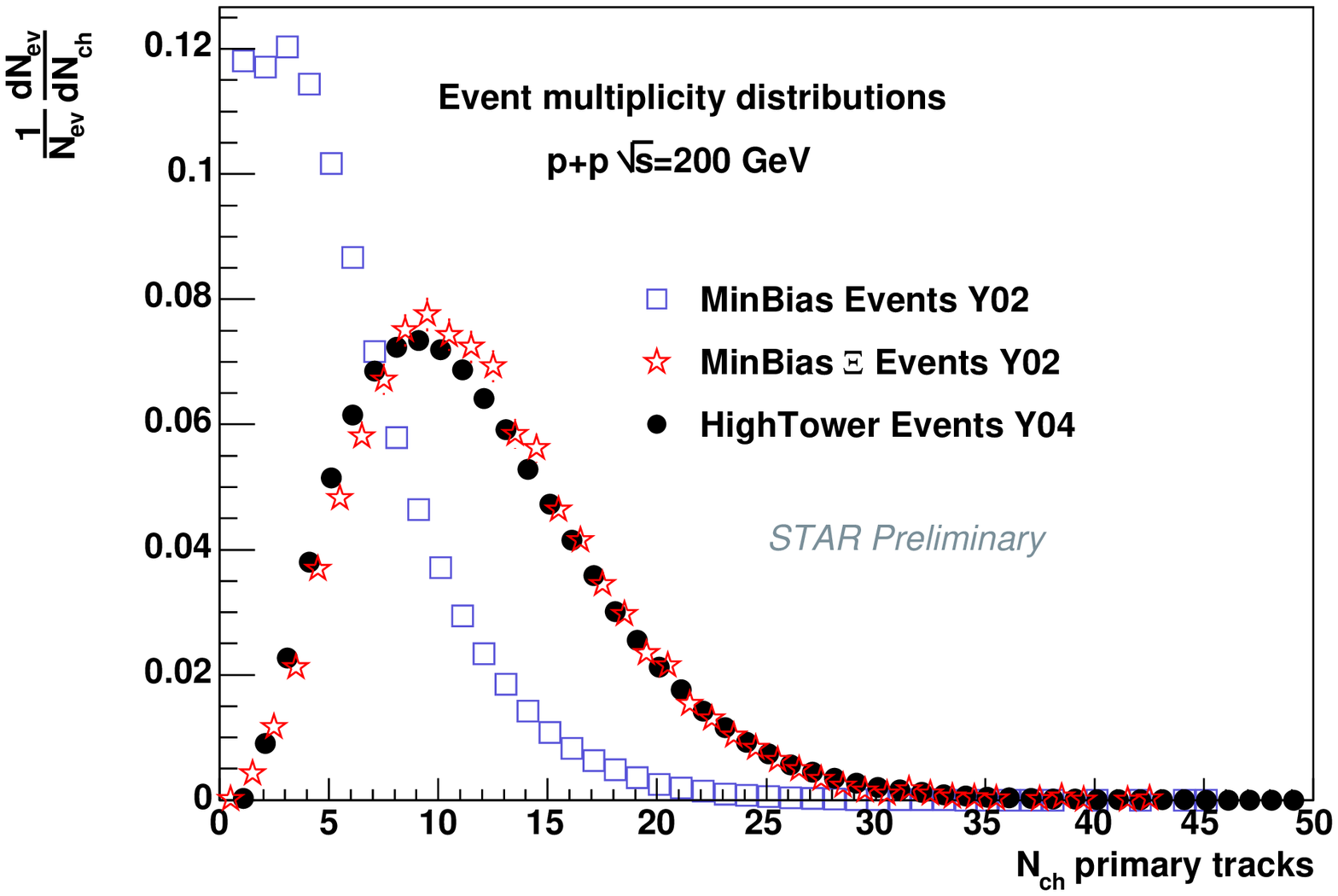}\\
 \vspace*{-1.2cm}\\
  \caption{Multiplicity distributions in all $\sqrt s=200$ GeV $p+p$ minimum bias events (squares),  minimum
  bias events with a $\Xi^-$ baryon (stars) and all triggered events (circles).}\label{fig4}
\end{minipage}
\end{figure}

By using triggered data, we increase the probability of producing
high $p_T$ $\Xi$ baryons suitable for correlations, as seen in
Fig.\ref{fig1}.  There we see raw $\Xi^-$ spectra normalized to
number of events in the respective data set. There are more $\Xi^-$
baryons produced in the triggered data than in the min. bias for
$p_T>2$ GeV/c, thus increasing the $\langle p_T\rangle$ of the
$\Xi^-$ spectrum. Therefore, triggering introduces a bias. To
understand this bias, we examine the difference between all min.
bias events and those min. bias events where we reconstruct a $\Xi$
baryon. We also compare multiplicity distributions of the two data
sets.
In the min. bias sample, as the multiplicity of the $p+p$ collision
increases, the $\langle p_T\rangle$ of produced particles has been
observed to rise \cite{opal}.  Fig.\ref{fig4} shows the difference
in uncorrected charged primary track distributions between min. bias
and high tower triggered events for $|y|<0.75$. The mean of the
former is $5.86\pm0.01$ charged tracks, while the mean of the latter
is $11.94\pm0.01$. When only min. bias events with a $\Xi$ baryon
are selected, the mean of the multiplicity distribution rises to
$12.12\pm0.05$. Thus events where a $\Xi$ baryon is produced belong
to the same multiplicity class as high-tower triggered events. Since
the enhanced multiplicity of the event might correspond to a higher
event $\langle p_T\rangle$ and thus to increased jettiness
\cite{gans}, we conclude that $\Xi$ baryons are likely to be
produced in jets.

Further evidence of a jet would be given by a same-side and a
back-to-back correlation between a $\Xi$ and the other particles in
the event. In order to construct such a correlation, a $\Xi^-$ with
$p_T>2$ GeV/c is used as a trigger particle, while its partner is
either a charged track ($h^{\pm}$ with $p_T>1.5$ GeV/c)
reconstructed in the TPC, or a charged or neutral particle detected
using an energy deposit in the EMC (the default cluster-finder
energy cut-off of 700 MeV was used). Then a correlation function is
constructed:

$$\displaystyle
C(\Delta\phi)=A_0e^{-\frac{\Delta\phi}{2\sigma^2_0}}+A_{\pi}e^{-\frac{
(\Delta\phi-\pi)}{2\sigma^2_{\pi}}}+B$$ where $A_0$ and $A_{\pi}$
are amplitudes of two Gaussian distributions of width $\sigma_0$ and
$\sigma_{\pi}$ respectively, centered at 0 and $\pi$ radians. In
$p+p$ collisions the Gaussians are assumed to sit on a flat
background of height $B$.  $\Delta\phi$ is calculated by measuring
the relative angle between the trigger particle and an associated
particle: a charged or a neutral hadron. Tracks and points that are
due to $\Xi$ decay products are excluded from the correlations. To
improve the statistics, after a raw correlation is obtained  the
data is "folded" around maxima at 0 and $\pi$ radians. The "folding"
is possible due to the expected symmetry of the away- and the same-
side peaks.

\begin{figure}[t!]
\begin{minipage}[t!]{2.3 in}
\vspace{-0.15cm}
  \includegraphics[width=2in]{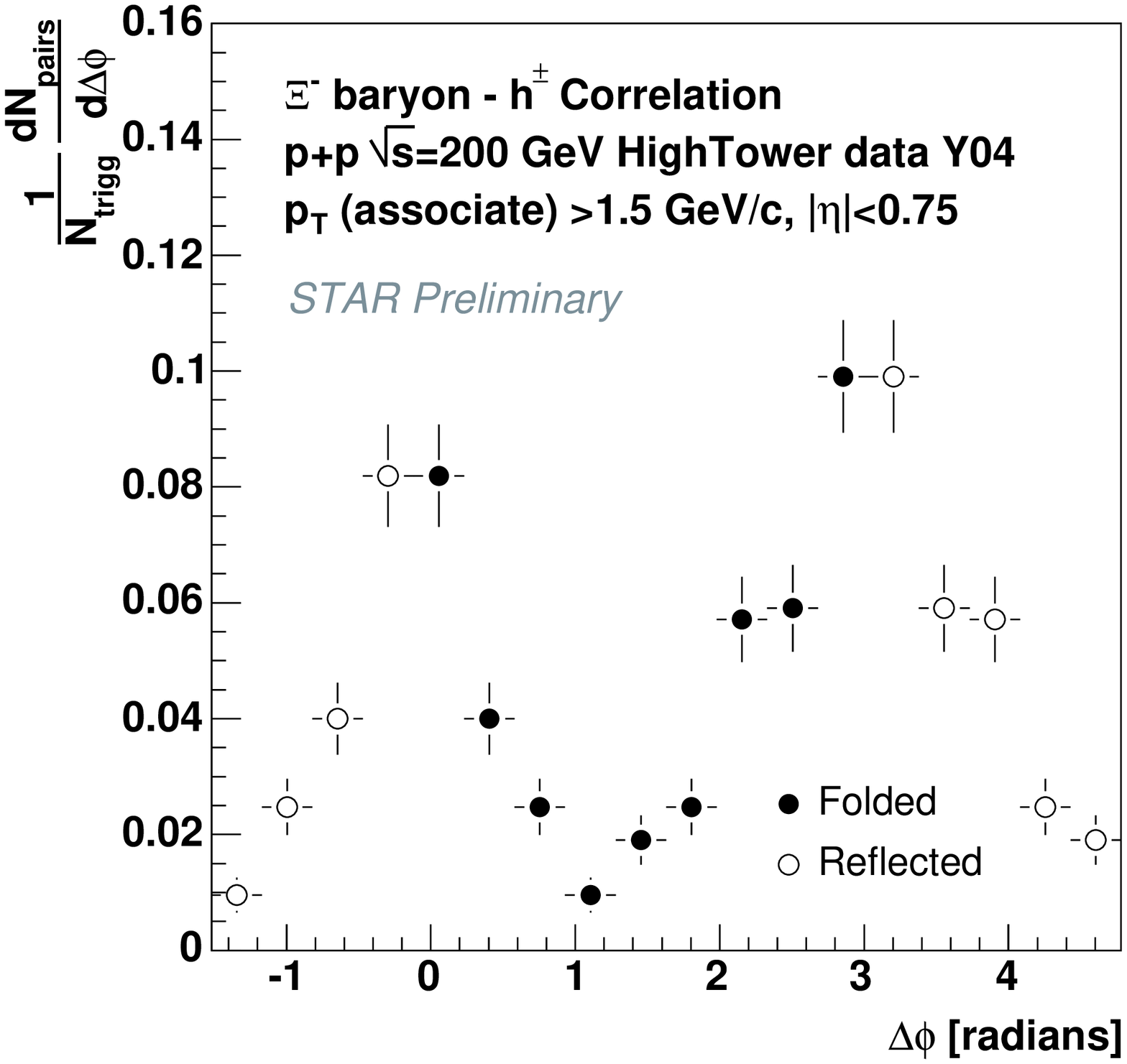}\\
  \vspace*{-1.5cm}\\
  \caption{Normalized $\Xi^--h^{\pm}$ correlations in triggered $\sqrt s=200$ GeV $p+p$ data.
  The points have been folded around 0 and $\pi$ radians.}\label{fig6}

\end{minipage}
\hfill
\begin{minipage}[]{2.3 in}
\vspace{-0.15 cm}
    \includegraphics[width=2 in]{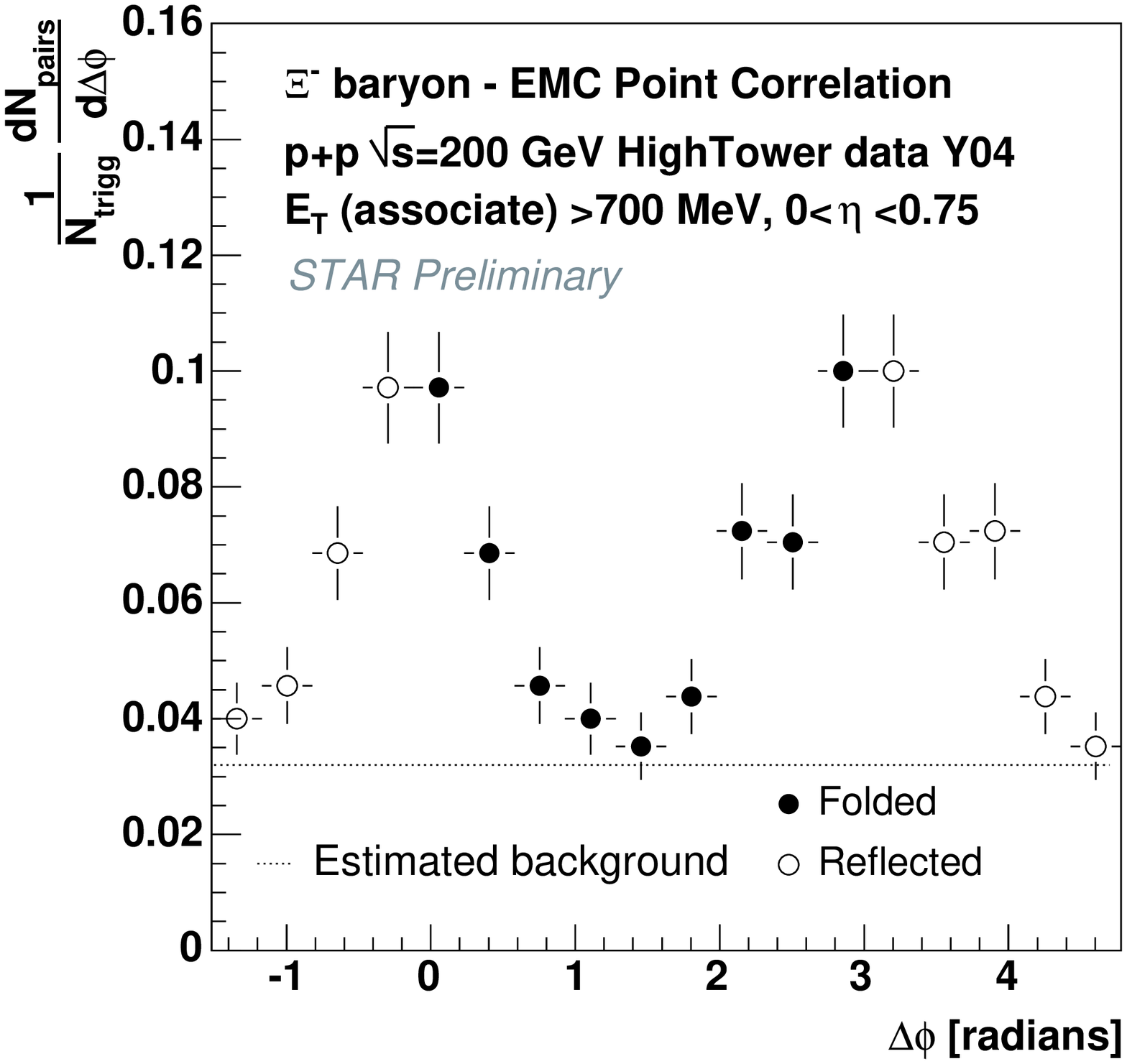}\\
\vspace*{-1.5cm}\\
  \caption{Normalized $\Xi^-$-EMC Point correlations in triggered $\sqrt s=200$ GeV $p+p$ data.
  The points have been folded around 0 and $\pi$ radians.}\label{fig5}
\end{minipage}
\end{figure}

While in the min. bias data set the statistics are poor
\cite{bbHQ04}, we see correlation functions for both $\Xi^--h^{\pm}$
(Fig.\ref{fig6}) and $\Xi$-EMC points (Fig.\ref{fig5}) in the
high-energy triggered data. The cut-off for an EMC point is set at
700 MeV.  Due to this low cut-off energy, the $\Xi$-EMC point
correlation function sits on a higher soft-particle correlation
background. The rapidity range for $h^{\pm}$ is $|y|<0.75$, whereas
it is $0<y<1$ for the EMC points. Because the tower points are not
identified at this stage of analysis, and the geometry as well as
efficiency of the correlations differ, a direct comparison between
$\Xi$-EMC point correlation yields and $\Xi-h^\pm$ yields cannot yet
be made.

\section{Conclusions}\label{concl}
Angular correlations between produced high $p_T$ $\Xi^-$ baryons and
unidentified charged and neutral particles (EMC points) are seen in
the high-tower triggered $p+p$ collisions. We see unambiguous
same-side and away-side peaks for both types of correlations.  The
high tower trigger seems to favor events with higher than average
multiplicity, and thus selects events where a $\Xi^-$ is more likely
to be produced than in the min. bias collisions.

Further comparison of high tower data to the min. bias sample is
needed before the results can be compared to those in Au+Au
collisions, or corrected yields calculated.  Once the bias is
determined, correlation widths and yields will allow
characterization of jets in which $\Xi^-$ candidates appear to be
produced. These yields and widths will then be compared to those in
the Au+Au data set. 70M Au+Au min. bias event are currently being
reconstructed for analysis. $\Xi$ correlation yields in comparison
to $\Lambda^0$ and $K^0_s$ yields should be insightful in $p+p$ as
well as in Au+Au collisions.


\vfill\eject
\end{document}